\documentclass[10pt,journal,compsoc]{IEEEtran}

\usepackage{amsmath,amsfonts,amssymb}
\usepackage{bm}
\usepackage{bbm}
\usepackage{dsfont}
\usepackage[normalem]{ulem}
\usepackage{pifont}
\usepackage{xcolor}
\usepackage{textcomp}

\newcommand{\circnum}[1]{\scalebox{1.45}{\raisebox{-0.1ex}{\ding{#1}}}}

\usepackage{algorithm}
\usepackage{algpseudocode}

\usepackage{array}
\usepackage{graphicx}
\usepackage{subfigure}
\usepackage{textcomp}
\usepackage{stfloats}
\usepackage{url}
\usepackage{hyperref}
\hypersetup{
    colorlinks=true,
    linkcolor=blue,
    filecolor=magenta,      
    urlcolor=cyan,
    citecolor=blue,
}\usepackage{verbatim}
\usepackage[capitalize,noabbrev]{cleveref}

\usepackage{cite}
\usepackage{ragged2e}
\usepackage{booktabs}
\usepackage{multirow}
\usepackage{makecell}
\usepackage{threeparttable}
\usepackage{longtable}
\usepackage{xtab}
\usepackage{lineno}
\usepackage{multicol}
\usepackage{tikz}
\usetikzlibrary{decorations.pathmorphing}
\usepackage[edges]{forest}

\usepackage{xcolor}
\usepackage{enumitem}
\usepackage{xspace}
\usepackage{diagbox}
\usepackage{pifont}
\usepackage{tcolorbox}

\newcommand*\circled[1]{\tikz[baseline=(char.base)]{\node[shape=circle, draw, inner sep=1.2pt] (char) {#1};}}

\newcommand{\summary}[1]{
    \begin{center}
    \begin{tcolorbox}[colback=gray!15, colframe=black, boxsep=-0.15cm, middle=-0.15cm]
        \textbf{\ding{46} Summary}
        $\blacktriangleright$
        {#1}
    $\blacktriangleleft$
    \end{tcolorbox}
    \end{center}
}

\newcommand{\ours}{\textsc{CodeDoc}\xspace}

\begin{document}

\title{Standards-Guided and Dependency-Aware Code-Level Documentation Generation}

\author{Tingxu~Han$^*$,
        Wei~Song$^*$,
        Ziming~Li$^*$,
        Yuchen~Chen,
        Weisong~Sun$^\dagger$,
        \\Zhenpeng~Chen,
        Chunrong~Fang$^\dagger$,
        Zhenyu~Chen,
        and~Yang~Liu
\thanks{Tingxu Han, Ziming Li, Yuchen Chen, Chunrong Fang, and Zhenyu Chen are with the State Key Laboratory for Novel Software Technology, Nanjing University, Nanjing, Jiangsu, 210093, China (e-mail: txhan@smail.nju.edu.cn; dzycd53@gmail.com; fangchunrong@nju.edu.cn; zychen@nju.edu.cn).}%
\thanks{Wei Song is with the School of Computer Science and Engineering, UNSW Sydney, NSW 2052, Australia (e-mail: wei.song1@unsw.edu.au).}%
\thanks{Weisong Sun and Yang Liu are with the College of Computing and Data Science, Nanyang Technological University, 639798, Singapore (e-mail: weisong.sun@ntu.edu.sg; yangliu@ntu.edu.sg).}%
\thanks{Zhenpeng Chen is with Tsinghua University, Beijing, China. E-mail: zpchen@tsinghua.edu.cn.}
\thanks{Tingxu Han, Wei Song, and Ziming Li contributed equally to this work.}%
\thanks{Weisong Sun and Chunrong Fang are corresponding authors.}%
}

\markboth{}%
{Han \MakeLowercase{\textit{et al.}}: Standards-Guided and Dependency-Aware Repository-Level Code Documentation Generation}

\maketitle

\begin{abstract}
Automatically generating high-quality documentation for software repositories remains challenging because existing approaches largely treat documentation as an unconstrained text generation problem.
As a result, they often lack explicit quality specifications, depend heavily on long-context prompts over raw code, and provide limited support for reproducible evaluation.
We argue that \textbf{code-level documentation generation} should instead be formulated as a \textbf{standards-guided software engineering task}, in which documentation is generated and assessed against explicit, reusable quality requirements derived from authoritative documentation standards.
In this paper, code documentation refers to natural-language descriptions of repository functions and classes, organized by source-code path. 

We present \ours, a standards-guided and dependency-aware framework for code-level documentation generation.
\ours identifies dependencies between code entities using static call-graph and AST analysis. It then constructs dependency-aware structured prompts that combine local code context, repository paths, and documentation criteria derived from authoritative standards and style guides.
We further introduce an interactive LLM-as-a-judge protocol that provides criterion-aligned feedback to optimize generated documentation, enabling targeted refinement based on fine-grained dimensions derived from formal and practice-oriented documentation standards.
Experiments on ten open-source Python repositories show that \ours produces 
higher-quality documentation than both RepoAgent and a long-context baseline 
under standards-aligned evaluation, with relative improvements of 
11.6\%--17.0\% and 21.4\%--29.6\%, respectively, while consuming roughly an 
order of magnitude fewer tokens per documentation entity than the 
long-context baseline.
A human evaluation study confirms substantial alignment between LLM-based 
and human judgments (Spearman's $\rho = 0.72$), and the relative findings 
are preserved across different judge LLMs.
Code is available at \url{https://github.com/ControllableModel/CodeDoc}. 
\end{abstract}


\section{Introduction}
\label{sec:intro}
Throughout the software lifecycle, developers continuously interact with existing code artifacts~\cite{pietri2020determining,ruparelia2010software,georgiou2019software,shafiq2021literature}, especially during tasks such as feature extension~\cite{van2025life}, bug fixing~\cite{zhong2015empirical}, code review~\cite{wang2021can}, and system maintenance~\cite{rajlich2014software}.
A growing body of empirical evidence shows that a substantial portion of developers' effort is devoted to understanding existing code rather than writing new functionality~\cite{oliveira2024understanding,oliveira2021recommending,rai2022review}.
In this context, high-quality code documentation plays a critical role in facilitating code comprehension, supporting collaboration, and improving the maintainability and evolution of software systems.
However, maintaining documentation manually is labor-intensive and error-prone~\cite{alzahrani2024software,shin2021survey}.
As software evolves, documentation often becomes incomplete, inconsistent, or outdated, reducing its usefulness and potentially harming overall software quality.
These challenges have motivated increasing research interest in automatic code documentation generation, with the goal of reducing manual effort while improving the availability and quality of software documentation~\cite{zhu2025reposummary,khan2022automatic,luo2024repoagent}.

\begin{figure}[!h]
    \centering
    \includegraphics[width=0.89\linewidth]{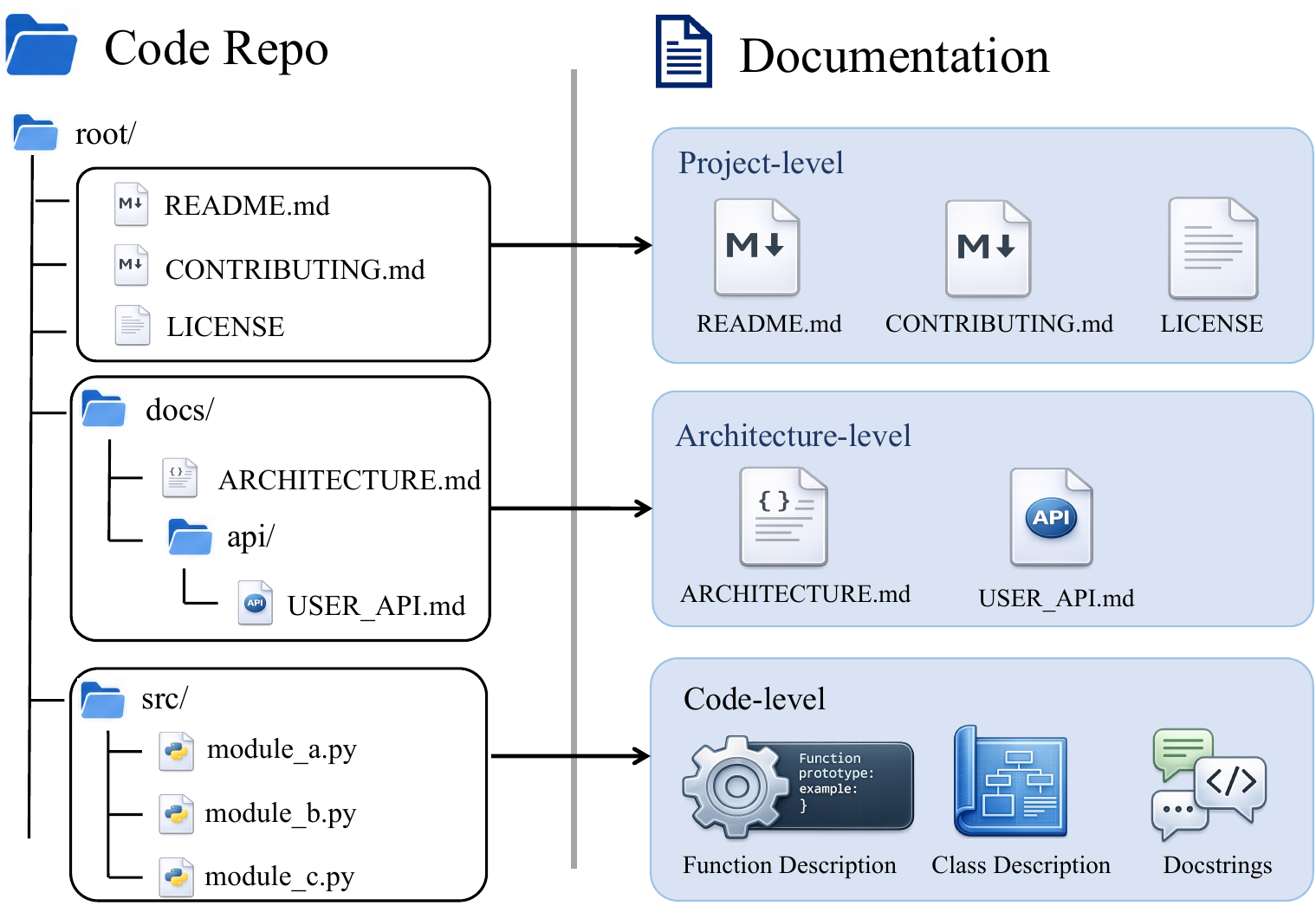}
    \caption{An intuitive example for different documentation levels for a code repo. We focus on code-level documentation in this paper.}
    \label{fig:code_doc}
\end{figure}

As shown in \autoref{fig:code_doc}, following~\cite{IEEE8292008}, documentation in a code repository can be organized at different levels. Project-level documentation describes the repository as a whole, such as README, CONTRIBUTING, and LICENSE files. Architecture-level documentation explains higher-level design and interface information, such as architecture descriptions and API documents. In contrast, this work focuses on code-level documentation, which refers to natural-language descriptions attached to individual code entities, such as function and class descriptions and docstrings.
Different from inline code comments, which are implementation-level annotations embedded in the code body that explain how or why particular statements work, code-level documentation describes an entity's purpose, behavior, interface, and role within the repository, forming a self-contained description.
Repository-level documentation can be constructed by organizing these entity-level descriptions according to the source code structure, but our study specifically targets generating documentation for a single code entity at a time.

Existing research on automatic code documentation generation falls into three strains. \emph{Heuristic-based} methods generate documentation through handcrafted templates, rules, or retrieval over similar fragments~\cite{abid2015using,mcburney2014automatic,rai2017method,haiduc2010use,haiduc2010supporting,vassallo2014codes,wong2015clocom,iyer2016summarizing}. They are interpretable and controllable, but rely on manually designed heuristics or code-similarity assumptions and generalize poorly to large, evolving codebases, especially in repository-level settings with cross-file and cross-function dependencies. \emph{Neural model-based} methods learn mappings from code to natural language using pretrained code models such as CodeNN~\cite{iyer2016summarizing}, CodeBERT~\cite{feng2020codebert}, and PLBART~\cite{ahmad2021unified}, substantially improving semantic modeling, but most focus on function-level summarisation rather than repository-wide coverage. \emph{LLM-based} methods leverage in-context learning and long-context capabilities of general-purpose LLMs to produce more fluent and adaptable documentation, and recent repository-level systems such as RepoAgent~\cite{luo2024repoagent} and RepoSummary~\cite{zhu2025reposummary} explicitly incorporate directory structures, code dependencies, and feature-level organization to improve global coherence beyond isolated method-level summaries.
Despite these advances, existing approaches still face several fundamental challenges:
\circled{1} They lack explicit and verifiable documentation criteria, leaving the notion of what constitutes high-quality documentation largely implicit and inconsistent across tools and usage scenarios;
\circled{2} They often incur high token costs because they rely on deep recursive analysis and redundant dependency inputs, which limit scalability and hinder practical deployment on large repositories.
\circled{3} Their code understanding can become unstable when excessively long contexts are provided, as critical information may be diluted by irrelevant details, leading to inconsistent or inaccurate documentation generation.
\circled{4} Their evaluation methods still rely heavily on human judgments and subjective assessments, which makes results harder to reproduce and limits objective comparison across methods.

To address the above challenges, we formulate automatic code documentation generation as a \textbf{standards-guided} and verifiable generation task, in which the content, structure, and scope of the generated documentation are explicitly constrained by well-defined documentation requirements derived from documentation standards.
In this work, we propose \ours, a standards-guided and dependency-aware framework for code-level documentation generation.
We first extract documentation standards from authoritative sources, including ISO/IEC/IEEE standards and widely used style guides, and convert them into explicit criteria that define what to document and how to structure it.
Guided by these criteria, \ours generates accurate and structured documentation by combining standards grounding with static program analysis.  It further incorporates dependency-aware context construction to capture relationships between code entities across the repository, and closes the loop with a criterion-aligned LLM judge whose feedback iteratively refines the generated documentation against the same standards.
The main contributions of this work are as follows:
\begin{itemize}[leftmargin=*, topsep=0pt]
    \item We present a standards-guided perspective on automatic code documentation generation, systematically deriving explicit and actionable documentation criteria from authoritative ISO/IEC/IEEE and GitHub documentation standards. The resulting criteria provide a reusable foundation for both defining and evaluating documentation quality.

    \item We propose \ours, a standards-guided and dependency-aware framework for code-level documentation generation. \ours couples standards-derived criteria with AST-based structural analysis and static call-graph dependency modeling, constructing dependency-aware structured prompts that improve repository-level code understanding while reducing redundant context and token consumption.

    \item We introduce a criterion-aligned LLM-as-a-judge protocol that plays a dual role: it provides dimension-level feedback to iteratively optimize the generated documentation in a closed loop, and it doubles as a reproducible, standards-grounded evaluation paradigm—so that a single explicit specification governs both generation and assessment.

    \item We conduct experiments on ten open-source Python repositories, showing that \ours improves documentation quality by 11.6\%–17.0\% over RepoAgent and 21.4\%–29.6\% over a long-context baseline under standards-aligned evaluation, while consuming roughly an order of magnitude fewer tokens per documentation entity. The LLM-based evaluation aligns with human judgment (Spearman's $\rho = 0.72$), and the relative findings are preserved across different judge model families.
\end{itemize}

\section{Background}
\label{background}

\subsection{Code-Level Documentation Generation}
\label{subsec:code_doc_repo_level}

Software documentation can describe a repository at different levels, from project-level files such as READMEs to architecture-level documents and code-level descriptions. 
ISO/IEC/IEEE standards~\cite{iso12207} treat source code as a distinct life-cycle artifact subject to inspection, analysis, verification, and validation. Following this source-code-level perspective, we define code-level documentation as natural-language descriptions attached to individual code entities, such as functions, methods, and classes. Typical examples include Python docstrings, Javadoc comments, and Doxygen documentation blocks, which are written alongside source code and can be extracted into API or reference documentation~\cite{pep257,javadoc,doxygen}. Code-level documentation generation therefore aims to automatically produce such entity-level descriptions from code, a task commonly studied as function or method documentation generation in prior work~\cite{barone2017parallel,feng2020codebert}. In this paper, we focus on generating documentation for a single code entity at a time, rather than producing project- or architecture-level documents.

\subsection{Formalizing Code-Level Documentation Generation as a Standards-Guided Task}
\label{subsec:code_doc_criterion_driven}
Let a code repository be represented as $\mathcal{C} = (\mathcal{E}, \mathcal{R})$, where $\mathcal{E}$ denotes a set of code entities (functions and classes within scope, plus enclosing modules used for path metadata), and $\mathcal{R}$ denotes the structural relations among them, such as call relations, import dependencies, and module hierarchies. The documentation target is \emph{code-level}: we produce one natural-language artifact per code entity, with the repository structure $\mathcal{R}$ serving as generation context rather than as the documentation unit. Formally, code-level documentation generation produces a set of per-entity artifacts $\mathcal{D} = \{d_e \mid e \in \mathcal{E}\}$, where each $d_e$ explains entity $e$ at an appropriate level of abstraction given its relevant context in $\mathcal{R}$.
Unlike unconstrained natural language generation, the quality of each documentation artifact is governed by a set of explicit documentation criteria $\mathcal{Q} = \{q_1, q_2, \ldots, q_n\}$ that are themselves derived from authoritative documentation standards (e.g., ISO/IEC/IEEE standards, mature open-source style guides). Each criterion $q_i$ specifies a concrete, assessable requirement, such as covering core functionality, clarifying inter-module dependencies, or describing usage assumptions. Collectively, these criteria define both the information that should be documented and how it should be organized. Under this formulation, code-level documentation generation can be modelled as a \emph{standards-guided generation task} in which the criterion set $\mathcal{Q}$ acts as the bridge between high-level documentation standards and low-level generation behaviour.
Because documentation quality is inherently graded rather than binary, we associate each criterion $q_i$ with a per-entity scoring function $s_i(d_e, \mathcal{C}) \in \{1, \ldots, S\}$ that measures how well artifact $d_e$ satisfies $q_i$, where $S$ is the maximum score (in our experiments, $S = 5$). The generation objective is the mean over entities and criteria:
\[
\mathcal{D}^{*} = \arg\max_{\mathcal{D}} \; \frac{1}{|\mathcal{E}|\,|\mathcal{Q}|}\sum_{e \in \mathcal{E}}\sum_{q_i \in \mathcal{Q}} s_i(d_e, \mathcal{C}).
\]
This matches our evaluation protocol, where the repository-level score is the mean of per-entity, per-dimension scores. Different instantiations of $\mathcal{Q}$ correspond to different documentation scenarios, giving a unified modelling framework for code-level documentation generation, and making criteria explicit and measurable supports principled generation and evaluation beyond implicit, prompt-driven approaches.
We note that the optimisation formulation above serves as a \emph{conceptual framework} that clarifies the generation objective rather than a problem that is solved directly. In practice, the method presented in \autoref{sec:meth} approximates this objective through iterative prompt-and-refine generation guided by criterion-level feedback, rather than through formal optimization over the documentation space.

\section{Methodology}
\label{sec:meth}

\begin{figure*}[!t]
    \centering
    \includegraphics[width=0.95\linewidth]{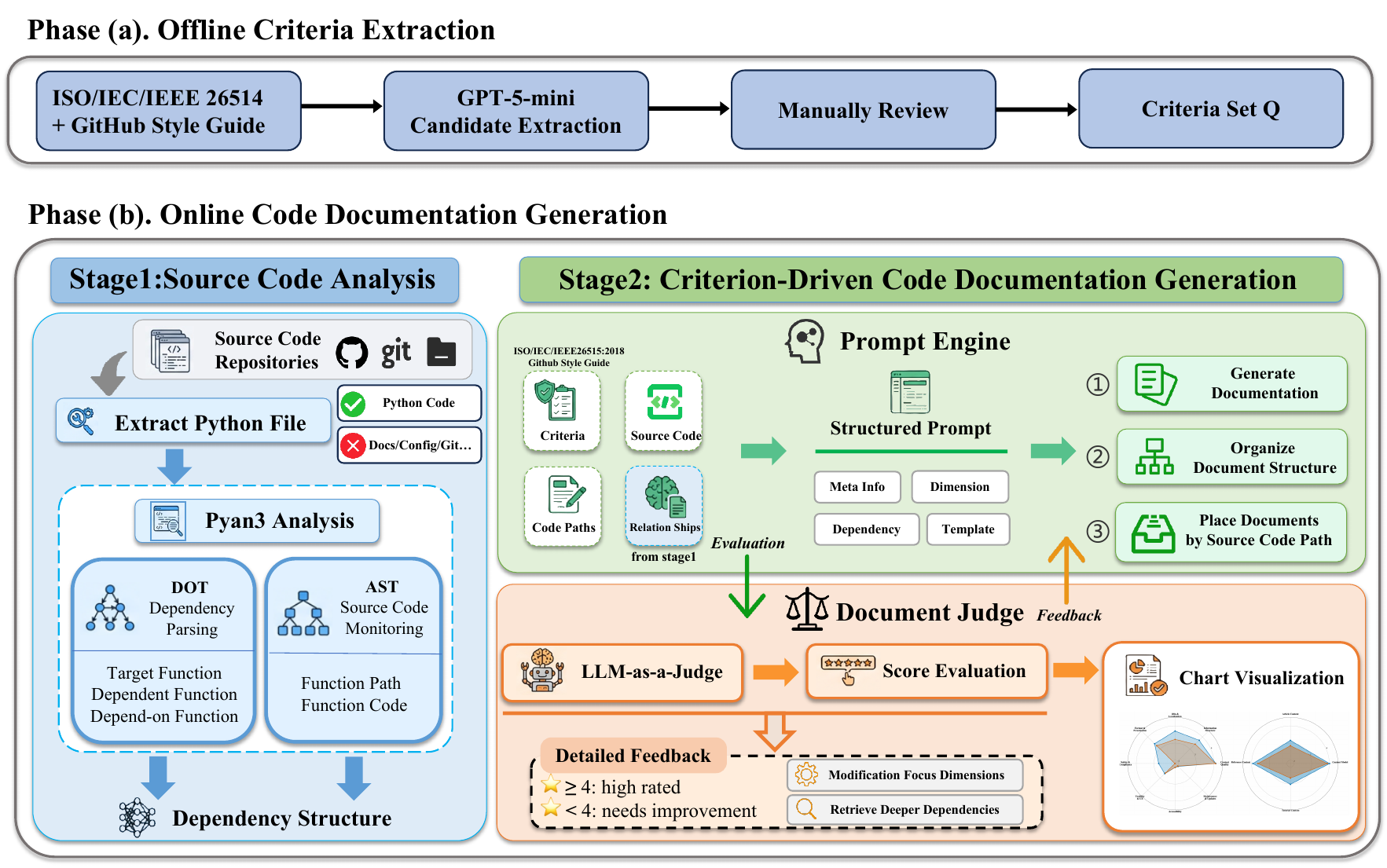}
    \caption{Overview of \ours{}, a standards-guided and dependency-aware framework for code-level documentation generation. \textbf{Phase~(a) Offline Criteria Extraction} (top) is run once, independently of any input repository: it turns authoritative documentation standards (ISO/IEC/IEEE~26514 and the GitHub style guide) into an explicit criterion set $\mathcal{Q}$ through GPT-5-mini candidate extraction followed by two-author manual review (\autoref{subsec:criteria_extraction}). \textbf{Phase~(b) Online Code Documentation Generation} (bottom) is the per-repository runtime pipeline. It comprises two stages. \emph{Stage~1} performs static source code analysis: it extracts function- and class-level entities from Python source files and constructs a dependency-aware structural representation via call-graph analysis (\texttt{Pyan3}) and AST parsing. \emph{Stage~2} feeds that representation together with $\mathcal{Q}$ into the prompt engine to produce standards-guided prompts for LLM-based documentation generation; the generated documents are organised by source-code path, and the LLM-as-a-Judge module scores them dimension by dimension and returns feedback for a single in-stage refinement pass.}
    \label{fig:framework}
\end{figure*}

In this paper, we present an automated framework for generating and evaluating code-level documentation by integrating static program analysis with large language models (LLMs).
As shown in \autoref{fig:framework}, the framework is organized into two phases.
\textbf{Phase~(a) Offline Criteria Extraction} (\autoref{subsec:criteria_extraction}) is performed once, ahead of any repository being processed, and turns authoritative documentation standards into an explicit, reusable criterion set $\mathcal{Q}$.
\textbf{Phase~(b) Online Code Documentation Generation} is the per-repository runtime pipeline; it consists of \emph{Stage~1} (\autoref{subsec:stage1}), which builds a dependency-aware structural representation from the source code, and \emph{Stage~2} (\autoref{subsec:stage2}), which combines that representation with $\mathcal{Q}$ to produce, organise, and refine the per-entity documentation.
Together, the three components support \textbf{standards-guided}, \textbf{dependency-aware}, and \textbf{token-efficient} code-level documentation generation, departing from prior work that relies on long-context prompting over raw source code without explicit quality specifications.

The remainder of this section describes the three components in order: \autoref{subsec:criteria_extraction} details Phase~(a); \autoref{subsec:stage1} and \autoref{subsec:stage2} detail Stages~1 and~2 of Phase~(b); and \autoref{subsec:walkthrough} walks through a representative example end-to-end.

\subsection{Phase~(a): Offline Criteria Extraction}
\label{subsec:criteria_extraction}

Both stages of Phase~(b) rely on an explicit set of documentation criteria $\mathcal{Q}$ derived once, offline, from authoritative documentation standards. The criteria simultaneously serve two roles: they constrain what Stage~2 must produce (as part of the structured prompt), and they form the rubric under which the document judge inside Stage~2, as well as any external evaluator (\autoref{sec:eval}), scores the generated artifacts. Making these criteria explicit, fine-grained, and standards-aligned is what distinguishes \ours{} from prior repository-level documentation systems that rely on implicit, ad-hoc prompts.

Source standards: we draw criteria from two complementary references that together cover formal and practice-oriented documentation expectations. \textbf{ISO/IEC/IEEE software documentation standards}~\cite{ISO} provide a formal engineering rubric covering correctness, completeness, consistency, modular organisation, maintainability, accessibility, internationalisation, safety, and update discipline; we adopt them as a baseline against which ``engineering-grade'' documentation is assessed. \textbf{The GitHub documentation style guide} encodes the practice of widely used open-source documentation, including developer-facing readability, task-oriented organization, navigability, procedural clarity, and contributor usability; we adopt it to complement the formal ISO rubric with the developer-experience axis.

Extraction pipeline: we operationalize the standards into machine-actionable criteria in two steps. (i)~\emph{Candidate extraction.} GPT-5-mini processes each source standard document and is prompted to identify distinct, assessable quality aspects, producing candidate items of the form \emph{(category, dimension, description, scoring guideline)} where each dimension corresponds to one independently verifiable property. This step yields an over-complete set that may contain duplicates and ambiguous phrasings. (ii)~\emph{Manual review and normalisation.} We manually inspect the candidates, deduplicate semantically overlapping entries, and normalize the wording so that each dimension maps to a concrete, independently scorable aspect.
The output is a fixed-schema rubric: each dimension carries a short name (used as the judge's variable), a one-sentence definition (used in the prompt), and a 1--5 scoring anchor (used to calibrate the judge).

Resulting rubrics: the final ISO rubric contains \textbf{7 categories} and \textbf{14 dimensions}; the final GitHub rubric contains \textbf{5 categories} and \textbf{12 dimensions}. \autoref{tab:iso_dimensions} and \autoref{tab:github_dimensions} list both rubrics with dimension definitions; a complete traceability table from source-standard clauses to extracted dimensions is provided in the supplementary material.

Usage in Phase~(b): the extracted criteria $\mathcal{Q}$ flow into Phase~(b) at two points: (i)~as the first block of every prompt assembled in Stage~2, so that the LLM generates documentation explicitly aligned with the rubric; and (ii)~as the scoring rubric used by the document judge inside Stage~2 as well as by external evaluators when reporting experimental results in \autoref{sec:eval}. This double use is precisely what makes the framework \emph{standards-guided}: a single explicit specification governs both generation and evaluation, rather than relying on separate, implicit notions of quality on either side.

\subsection{Phase~(b), Stage~1: Source Code Analysis}
\label{subsec:stage1}
Stage~1 builds the refined repository representation $\mathcal{C}'=(\mathcal{E},\mathcal{R}')$ that serves as the structural interface to Stage~2. Following the scope defined in \autoref{subsec:code_doc_repo_level}, the analysis is performed at function and class granularity: each function or class is recorded as an entity in $\mathcal{E}$ together with its repository path and source code body, preserving traceability for downstream documentation placement. Two complementary static analyses populate $\mathcal{R}'$. (i)~Call-graph analysis via \texttt{Pyan3} resolves imports and call sites to produce a repository-level directed call graph; the resulting caller--callee edges form the backbone of $\mathcal{R}'$. (ii)~AST parsing of each entity extracts localised structural cues---function signatures, parameter lists, class attributes, and major control-flow constructs---which complement the global call-graph structure with local semantic grounding. The integrated representation lets Stage~2 retrieve only dependency-relevant context, reducing redundant inputs and improving repository-level coherence under a controlled token budget.

\begin{figure*}[!t]
    \centering
    \includegraphics[width=0.85\linewidth]{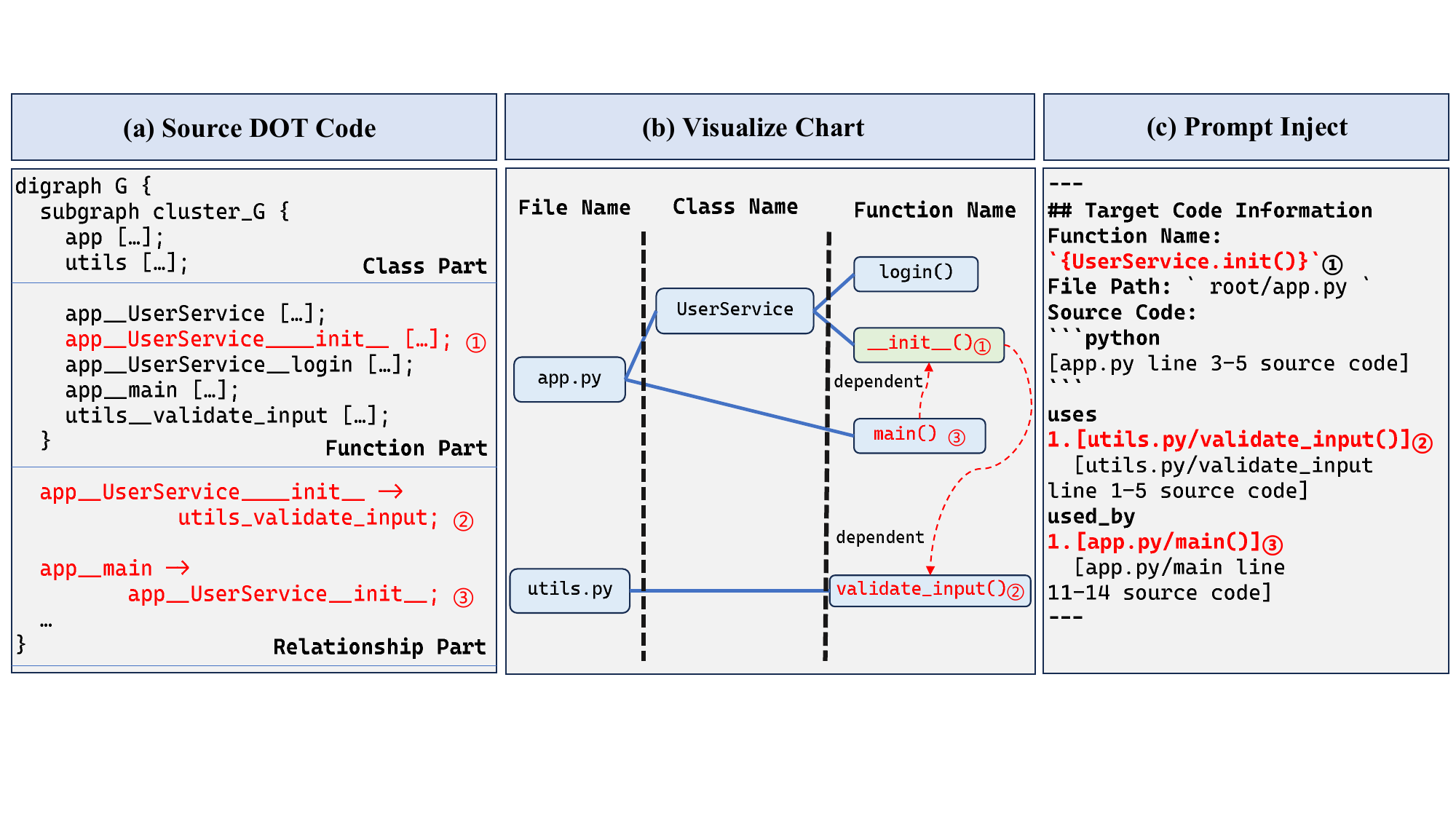}
    \vspace{-42pt}
    \caption{Dependency-graph construction and dependency-aware prompt injection in \ours{}. (a) The DOT code structure is organized into three sections: Class Part, Function Part, and Relationship Part. (b) The visualized DOT graph, which includes two files, one class, four functions, and two call relationships. (c) Illustrates how the information derived from the DOT structure is utilized in our prompt template---dependency relations are injected into the ``uses'' and ``used\_by'' sections with markers, ready for subsequent AST-based source code injection.}
    \label{fig:DOT}
    \vspace{6pt}
\end{figure*}

\begin{figure*}[!t]
    \centering
    \includegraphics[width=0.85\linewidth]{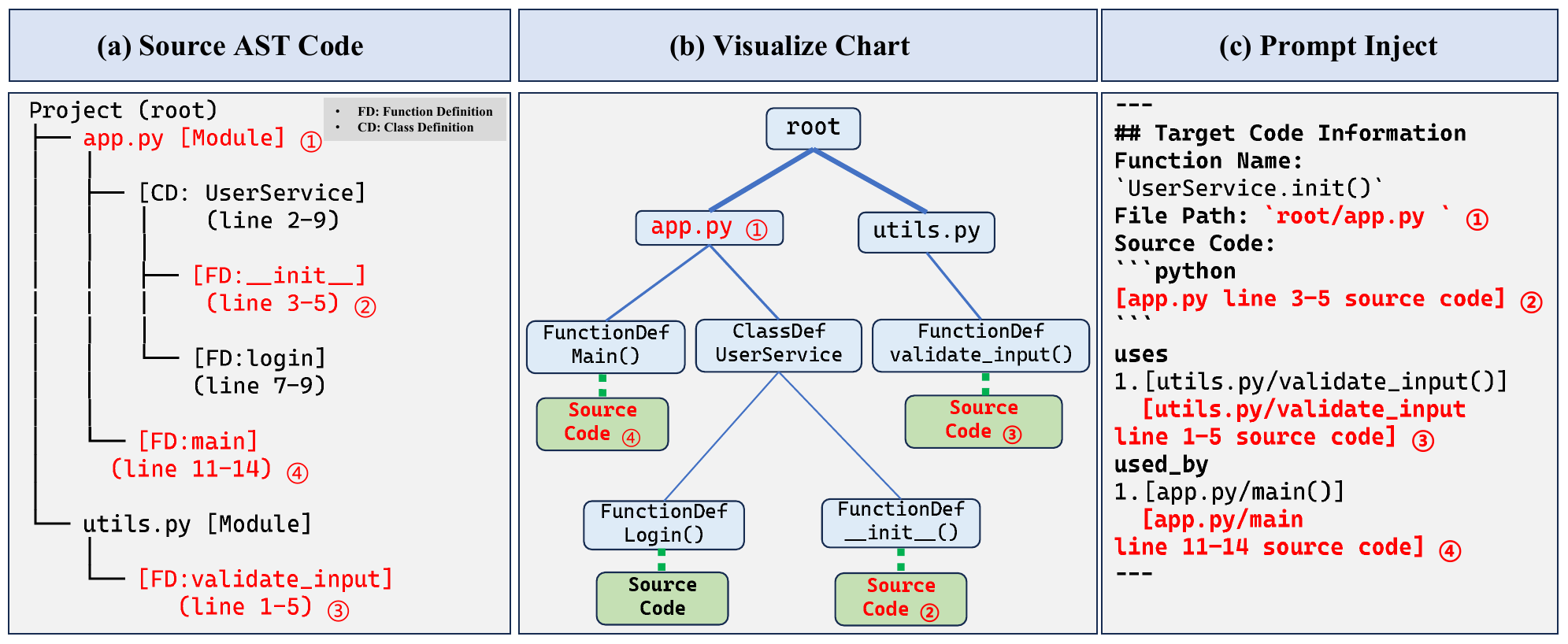}
    \caption{AST-based structural extraction and prompt construction in \ours{}. (a) The raw AST tree printed from the parsed repository, displaying the starting and ending positions of all classes and functions within each file starting from the root directory. (b) The visualised AST hierarchy, where each class or function node is annotated with its corresponding starting and ending line numbers in the source code. (c) Based on the DOT dependency structure, the AST injects the source code of retrieved functions into the corresponding ``uses'' and ``used\_by'' positions, forming a comprehensive prompt template.}
    \label{fig:AST}
\end{figure*}

\subsection{Phase~(b), Stage~2: Standards-Guided Documentation Generation}
\label{subsec:stage2}
Stage~2 takes the refined representation $\mathcal{C}'$ together with the criterion set $\mathcal{Q}$ from Phase~(a), and produces per-entity documentation through an iterative prompt-and-refine loop (see \autoref{fig:framework}). For each target entity the prompt engine selectively retrieves a small set of dependency-relevant entities from $\mathcal{R}'$---typically the direct callers and callees---rather than exposing the LLM to the entire repository. It then assembles a structured prompt that combines four blocks in a fixed order: (i)~documentation criteria definitions drawn from $\mathcal{Q}$, (ii)~target entity source code with its repository path, (iii)~dependency-relevant entity source code with caller/callee annotations, and (iv)~generation instructions specifying the expected output format. The LLM produces documentation that is anchored in the entity's local implementation while remaining aware of its role within the broader dependency structure. The output is organised by source-code path so that each artifact remains traceable to its code entity.

Within the same stage, an LLM-based \emph{document judge} evaluates the generated documentation against the same criterion set $\mathcal{Q}$, producing dimension-level scores and textual feedback. The feedback is appended to the original prompt for a single regeneration pass (we use one refinement iteration across all experiments); this closes the loop within Stage~2 without requiring repeated long-context analysis of the entire repository. All generation and evaluation use GPT-5-mini with default temperature settings, and example prompt templates and representative outputs are provided in the supplementary material.

\subsection{Walk-through Example}
\label{subsec:walkthrough}

We now walk through Phase~(b) end-to-end on a concrete Python repository consisting of two files, \texttt{app.py} and \texttt{utils.py}, as illustrated in \autoref{fig:DOT} and \autoref{fig:AST}.

\textbf{Phase (a): Offline Criteria Extraction.} Before Phase~(b) begins, Phase~(a) runs offline to produce the criterion set $\mathcal{Q}$, extracted from authoritative documentation standards (the ISO/IEC/IEEE documentation standards and the GitHub documentation style guide). The same $\mathcal{Q}$ is used downstream by both the prompt engine and the document judge, so that a single explicit specification governs both generation and evaluation.

\textbf{Phase (b)-Stage 1: Source Code Analysis.} Stage~1 builds the refined representation $\mathcal{C}'=(\mathcal{E},\mathcal{R}')$ through two complementary static analyses. \emph{(i) Call-graph analysis.} \texttt{Pyan3} analyses the whole repository and emits a directed call graph in DOT format. \autoref{fig:DOT}(a) shows the raw DOT output for our running example, which declares entities such as \texttt{app\_\_UserService\_\_\_\_init\_\_}, \texttt{utils\_\_validate\_input}, and \texttt{app\_\_main}, together with caller--callee edges from \texttt{app\_\_main} to \texttt{app\_\_UserService\_\_init\_\_} and from \texttt{app\_\_UserService\_\_\_\_init\_\_} to \texttt{utils\_\_validate\_input}. Parsing this DOT file yields the call graph in \autoref{fig:DOT}(b), whose nodes are the code entities in $\mathcal{E}$ and whose caller--callee edges form the backbone of $\mathcal{R}'$. At this point the representation captures only dependency structure; no source code is attached, since the DOT format records relations without implementation details. \emph{(ii) AST parsing.} AST parsing then complements the call graph with localised structural cues. \autoref{fig:AST}(a) presents the AST hierarchy of the same repository: the root node represents the project, with child modules \texttt{app.py} and \texttt{utils.py}; \texttt{app.py} contains the class \texttt{UserService} (lines 2--9), which encloses the methods \texttt{\_\_init\_\_} (lines 3--5) and \texttt{login} (lines 7--9), together with the top-level function \texttt{main} (lines 11--14); \texttt{utils.py} contains the standalone function \texttt{validate\_input} (lines 1--5). From this tree the analysis extracts function and class signatures, parameter lists, and precise line ranges for each entity, and uses the recorded line ranges to attach the corresponding source code to each node. \autoref{fig:AST}(b) visualises the resulting refined representation $\mathcal{C}'=(\mathcal{E},\mathcal{R}')$, in which each entity now carries both its caller--callee relations and its exact source location.

\textbf{Phase (b)-Stage 2: Standards-Guided Documentation Generation.} Stage~2 takes $\mathcal{C}'$ and the criterion set $\mathcal{Q}$ and assembles the structured prompt. For the target entity \texttt{UserService.\_\_init\_\_}, the prompt engine selectively retrieves its direct callers and callees from $\mathcal{R}'$: it identifies \texttt{main()} as a caller and \texttt{validate\_input()} as a callee and records them under the \texttt{used\_by} and \texttt{uses} sections of the prompt, respectively (\autoref{fig:DOT}(c)). It then fills these sections with the source code retrieved via the AST line ranges---\texttt{validate\_input()} from \texttt{utils.py} and \texttt{main()} from \texttt{app.py}---alongside the target entity's own file path and source code (\autoref{fig:AST}(c)). Combined with the repository-path metadata and the criteria $\mathcal{Q}$, these form the structured prompt, from which the LLM generates documentation that is anchored in the target entity's local implementation while remaining aware of its role in the broader dependency structure.
Finally, the document judge applies $\mathcal{Q}$ to score the generated documentation and produce textual feedback, which the prompt engine consumes for a single refinement pass before the LLM regenerates the documentation. This closes the loop within Stage~2 without requiring repeated long-context analysis of the entire repository.

\section{Experimental Evaluation}
\label{sec:eval}

This section evaluates \ours{} along a deliberate progression: we first establish that our measuring instrument is sound, then use it to assess quality, and finally dissect the cost and mechanism behind the results. RQ1 asks whether the standards-derived, LLM-based evaluation is itself trustworthy---sensitive to quality degradation, aligned with human judges, and stable across judge families. RQ2 then uses this validated protocol to test whether \ours{} improves documentation quality over baselines and whether the gain is broad-based. RQ3 finally asks at what token cost these gains come, which components drive them, and how reliable the static dependency recovery underpinning the method is. Each question builds on the previous: RQ2 relies on the rubric validated in RQ1, and RQ3 interprets the cost of the quality established in RQ2.
\begin{itemize}[leftmargin=*]
    \item \textbf{RQ1 (Evaluation Validity):} Is the standards-derived, LLM-based evaluation a trustworthy measure of documentation quality---i.e., is it sensitive to controlled quality degradation, aligned with human judgement, and stable across judge model families?
    \item \textbf{RQ2 (Documentation Quality):} Does \ours{} produce higher-quality documentation than existing baselines under the standards-aligned evaluation, and is the improvement broad-based across quality dimensions?
    \item \textbf{RQ3 (Cost, Components \& Reliability):} At what generation cost does \ours{} achieve these gains, which of its design choices drive them, and how reliable is the static dependency recovery the method depends on?
\end{itemize}

\begin{figure*}
    \centering
    \includegraphics[width=0.89\linewidth]{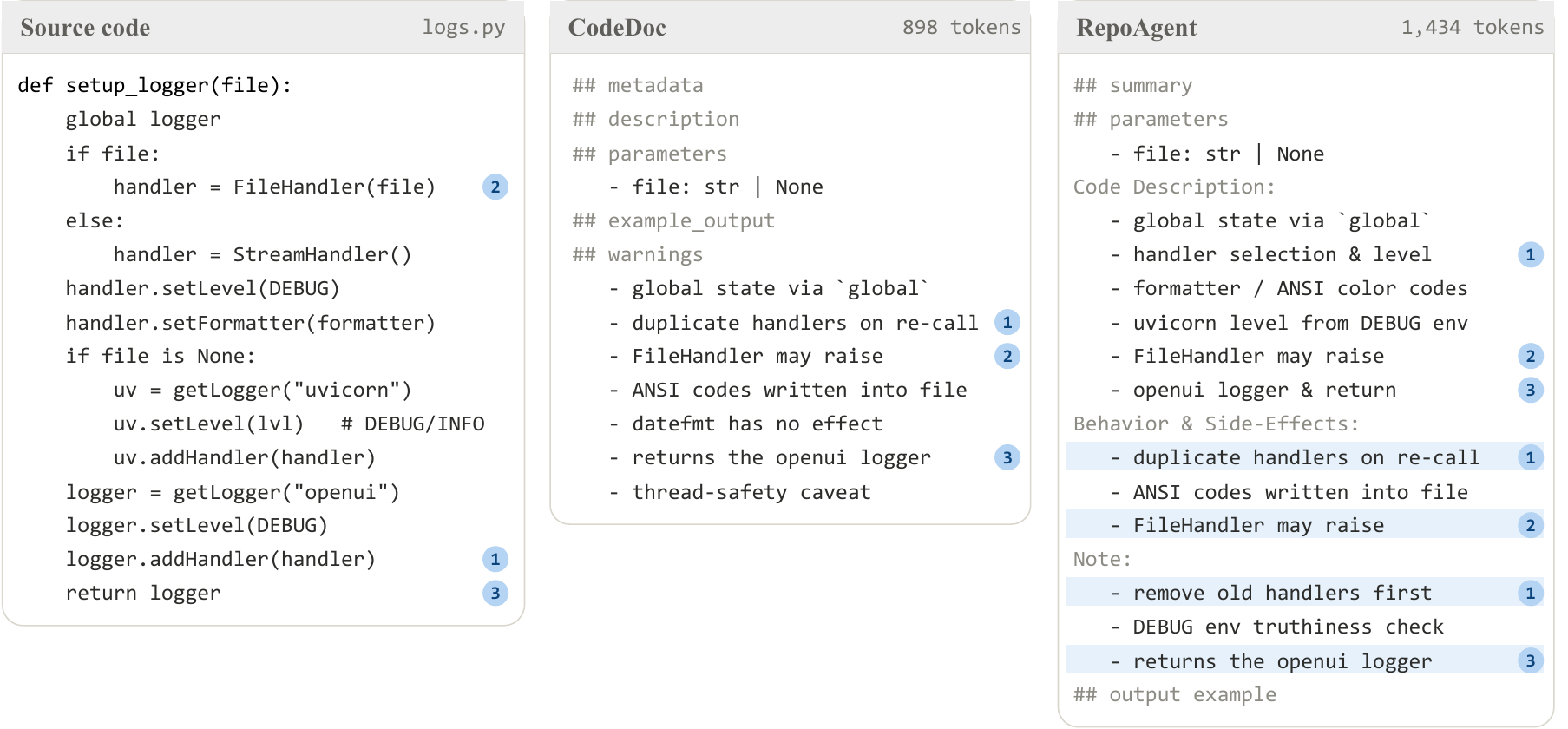}
    \caption{A qualitative comparison of documentation generated for the \texttt{setup\_logger} function (\texttt{openui/logs.py}) by \ours (898 tokens) and RepoAgent~\cite{luo2024repoagent} (1,434 tokens). Circled markers \circnum{172}--\circnum{174} trace individual documentation statements back to their corresponding source-code lines. \ours{} yields a more concise, criterion-structured artifact that still covers the key behaviors and side-effects, e.g., duplicate handlers on re-call \circnum{172}, a possible \texttt{FileHandler} exception \circnum{173}, and the returned \texttt{openui} logger \circnum{174}, at substantially lower token cost.}
    \label{fig:evidence_redun}
\end{figure*}

\subsection{Experimental Setup}
\label{subsec:setup}

Repositories: We evaluate ten open-source Python repositories spanning AI and software-infrastructure domains (\autoref{tab:repo_stats}). Each repository is preprocessed by removing directories unrelated to source code (datasets, build artifacts, configuration files, and pre-existing documentation) so that all methods are run on identical inputs. We measure repository scale by the total number of documentation entities (\#Class~+ \#Function) and group repositories into \emph{small} ($\leq 200$), \emph{medium} ($200{<}{\cdot}{\leq}400$), and \emph{large} ($>400$).

Evaluation rubric: All experiments evaluate documentation quality under the two standards-derived rubrics introduced in \autoref{subsec:criteria_extraction}: the ISO rubric (7 categories, 14 dimensions, drawn from ISO/IEC/IEEE software documentation standards~\cite{ISO}) and the GitHub rubric (5 categories, 12 dimensions, drawn from the GitHub documentation style guide). See \autoref{tab:iso_dimensions} and~\autoref{tab:github_dimensions} for the full rubrics.

LLM-based judge: Following recent work~\cite {zheng2023judging,liu2023geval}, we use an LLM judge to score generated documentation against predefined criteria. For each dimension, the judge assigns a $1$--$5$ score (higher is better) and is instructed to focus exclusively on the documentation artifact and the corresponding criterion. The repository-level score under a given standard is the mean of dimension-level scores. To examine whether the judge's interaction style affects evaluation stability, we report results under two scoring strategies: \emph{single-round} (S), where all dimensions are scored within a single-round conversation, and \emph{multi-round} (M), where each dimension is scored in an independent interaction

Statistical reporting: the unit of analysis is the repository ($n=10$): each repository yields one paired observation (baseline vs.\ \ours{}). For each pairwise comparison we report the mean score difference, Cliff's delta effect size, and the Wilcoxon signed-rank $p$-value across repositories. Given the small sample size, we interpret results based on the combination of consistent per-repository trends, score distributions, effect sizes, and statistical indicators rather than relying on any single metric.

\subsection{RQ1: Is the Standards-Based Evaluation Trustworthy?}
\label{subsec:rq1}

RQ1.a~--~Criterion sensitivity (controlled masking).
To answer whether the extracted criteria capture quality, we use a controlled-masking method: if the criteria genuinely capture quality, their scores should fall as quality is progressively removed. We therefore validate whether the extracted criteria respond sensibly to controlled quality degradation. Using official PyTorch~2.8 documentation files~\cite{pytorch_docs}, we select all files larger than 20\,KB as high-quality references and randomly mask $20\%$, $40\%$, or $60\%$ of each file's content to obtain four document groups (one clean, three progressively degraded). All groups are scored by the LLM judge under the multi-round strategy only (to isolate criterion sensitivity from intra-dialogue effects) under both ISO- and GitHub-based criteria.

\begin{figure*}[!t]
    \centering
    \includegraphics[width=0.85\linewidth]{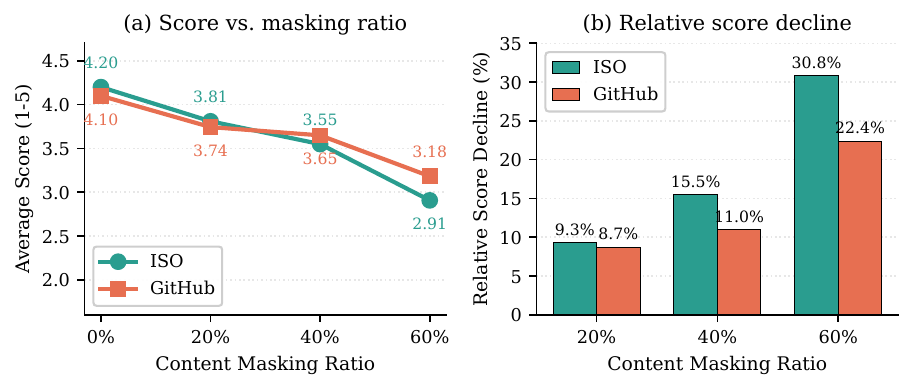}
    \caption{Responses of the extracted criteria to controlled documentation degradation. \textbf{(a)} Average score under progressive content masking (0--60\%). \textbf{(b)} Relative decline from the unmasked baseline. Both rubrics decline monotonically as the masking ratio grows, with ISO dropping faster than GitHub at high masking ratios.}
    \vspace{-8pt}
    \label{fig:rq1_sensitivity}
\end{figure*}

\autoref{fig:rq1_sensitivity} shows that both criterion sets assign progressively lower scores as the masking ratio increases. The monotonic decline indicates that the extracted dimensions respond to controlled information loss rather than producing arbitrary scores. The two standards differ in magnitude: the ISO criteria drop more sharply under heavy masking (sensitive to structural incompleteness), whereas the GitHub criteria degrade more gently (consistent with their emphasis on readability rather than formal completeness alone).

RQ1.b~--~Human alignment and cross-family judge robustness.
We validate the LLM-based evaluation along two axes: alignment with human judges, and robustness across judge model families. We sample three repositories spanning different scales---Flask (small), NumPy (medium), and PyTorch (large)---and randomly select 15 functions per repository, yielding 45 documentation pairs (one from \ours{} and one from RepoAgent per function). Three graduate students with two or more years of Python development experience score every artifact (blinded, randomised order) on a $1$--$5$ scale across a representative subset of eight ISO dimensions. To assess robustness across judge families, we re-evaluate the same 45 pairs using Claude-3.5-Sonnet under an identical rubric.

\autoref{tab:rq5_combined} reports both views. Human evaluators consistently prefer \ours{} ($+0.52$ on average; inter-rater Fleiss' $\kappa = 0.51$, moderate). The Spearman correlation between average human scores and GPT-5-mini judge scores across all 45 pairs is $\rho = 0.72$ ($p<0.001$); human scores are systematically $0.1$--$0.2$ points lower than LLM-judge scores, consistent with known mild LLM-judge inflation~\cite{zheng2023judging}. The relative ranking is fully preserved when switching the judge to Claude-3.5-Sonnet, with $\rho = 0.85$ between the two judges at the dimension level and a near-identical score gap between methods ($\Delta = 0.55$ vs.\ $0.52$).

\begin{table}[!t]
\centering
\caption{Human Alignment and Cross-Family Judge Comparison.}
\label{tab:rq5_combined}
\renewcommand{\arraystretch}{1.15}
\setlength{\tabcolsep}{5pt}
\footnotesize
\begin{tabular}{l|cc|cc|cc}
\toprule
\multirow{2}{*}{\textbf{Method}}
 & \multicolumn{2}{c|}{\textbf{Human}}
 & \multicolumn{2}{c|}{\textbf{GPT-5-mini}}
 & \multicolumn{2}{c}{\textbf{Claude-3.5-Sonnet}} \\
 & Score & Rank & Score & Rank & Score & Rank \\
\midrule
RepoAgent & 3.09 & 2 & 3.24 & 2 & 3.11 & 2 \\
\ours{}   & \textbf{3.61} & 1 & \textbf{3.79} & 1 & \textbf{3.63} & 1 \\
$\Delta$  & +0.52 & -- & +0.55 & -- & +0.52 & -- \\
\bottomrule
\end{tabular}
\end{table}

\summary{Both standards detect quality differences in a controlled masking test, declining monotonically with information loss. The LLM-based evaluation correlates strongly with human judgement ($\rho = 0.72$) and the relative method ranking is preserved across judge families, mitigating same-family bias.}

\subsection{RQ2: Does \ours{} Improve Documentation Quality?}
\label{subsec:rq2}

RQ2.a~--~Quality vs.\ baselines.
Having established that the criteria distinguish quality levels, we compare \ours{} against two baselines: \emph{RepoAgent}~\cite{luo2024repoagent}, the most directly comparable open-source repository-level documentation generation tool, and a \emph{Long-Context} baseline that feeds the full source file containing each target entity to the same backbone LLM (GPT-5-mini) under a matched token budget, without documentation criteria or dependency selection. For each repository we generate documentation under both ISO- and GitHub-oriented objectives, and evaluate each generation only under its aligned criteria.

\autoref{tab:rq2_extended} reports aggregate scores. \ours{} achieves the highest score in every cell, with margins of $+0.39$ to $+0.55$ over RepoAgent and $+0.67$ to $+0.86$ over Long-Context. All pairwise improvements are statistically significant (Wilcoxon $p<0.05$, Cliff's $\delta>0.6$ in every setting).

\begin{table}[!t]
\centering
\caption{Average Documentation Scores Across Ten Repositories Under the ISO and GitHub Standards. Single-Round (S) and Multi-Round (M) Evaluation. 
Best in \textbf{Bold}.}
\label{tab:rq2_extended}
\renewcommand{\arraystretch}{1.15}
\setlength{\tabcolsep}{6pt}
\footnotesize
\begin{tabular}{l|cc|cc}
\toprule
\multirow{2}{*}{\textbf{Method}} & \multicolumn{2}{c|}{\textbf{ISO}} & \multicolumn{2}{c}{\textbf{GitHub}} \\
& S & M & S & M \\
\midrule
Long-Context & 2.87 & 2.93 & 3.01 & 3.08 \\
RepoAgent    & 3.18 & 3.24 & 3.29 & 3.35 \\
\ours{}      & \textbf{3.72} & \textbf{3.79} & \textbf{3.68} & \textbf{3.74} \\
\bottomrule
\end{tabular}
\end{table}

\begin{figure*}[!t]
    \centering
    \includegraphics[width=0.85\linewidth]{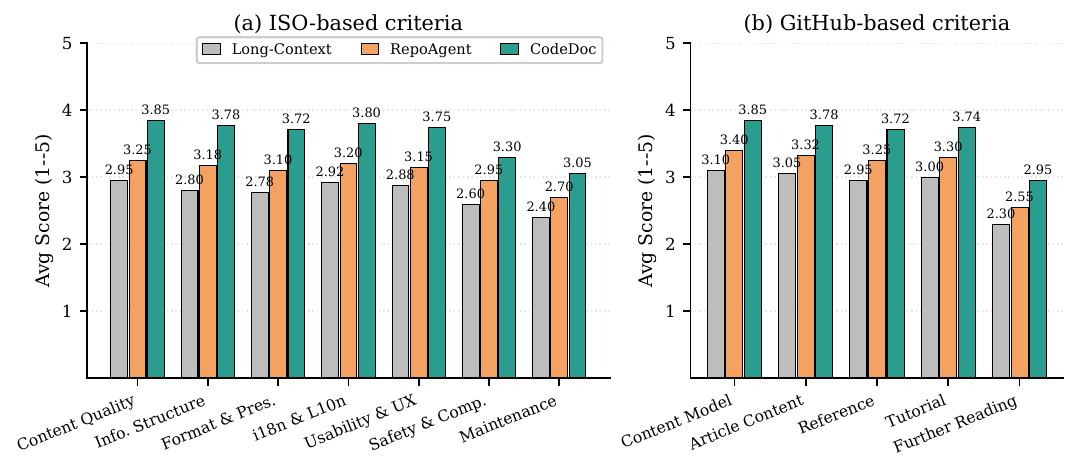}
    \caption{Per-category documentation-quality comparison between baselines and \ours{} under the ISO and GitHub criteria. Scores are averaged across the dimensions within each category.}
    \label{fig:rq1_quality_bars}
\end{figure*}

RQ2.b~--~Category-level breakdown.
A category-level decomposition (\autoref{fig:rq1_quality_bars}) confirms that the improvement is broad-based rather than driven by one or two dimensions: \ours{} leads in every ISO and GitHub category, with the largest absolute gains in \textit{Content Quality}, \textit{Information Structure}, and \textit{Format \& Presentation} for ISO, and \textit{Content Model} and \textit{Reference Content} for GitHub.

\summary{Across all ten repositories, \ours{} outperforms RepoAgent ($+0.39$ to $+0.55$) and Long-Context ($+0.67$ to $+0.86$) under both standards and both judge interaction strategies, with significant effect sizes and consistent category-level gains.}

\subsection{RQ3: Cost--Quality Trade-off, Component Contributions, and Mechanism Reliability}
\label{subsec:rq3}

RQ3.a~--~Cost--quality trade-off.
Higher documentation quality is valuable only at reasonable generation cost, so we examine token cost and documentation quality jointly. 

\begin{figure*}[!t]
    \centering
    \includegraphics[width=0.85\linewidth]{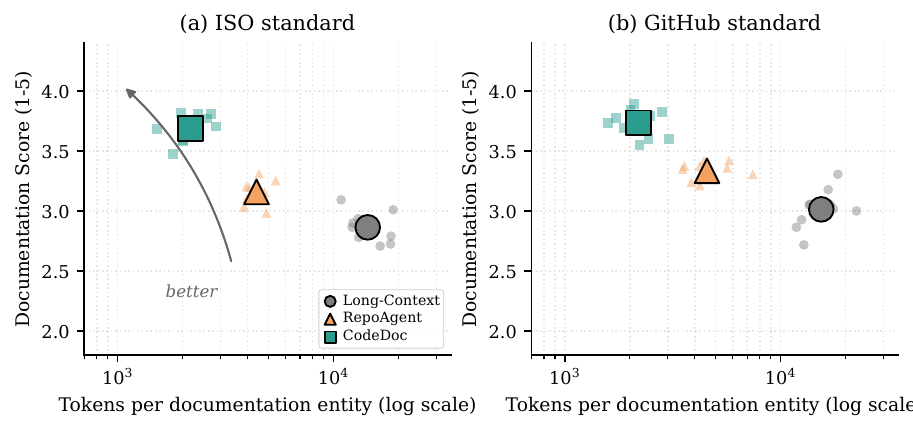}
    \caption{Cost--quality trade-off of \ours{} and the two baselines under (a) ISO and (b) GitHub criteria. Small markers are individual repositories; large markers are method-level centroids. Lower-left $=$ cheaper; upper-left $=$ better quality at lower cost.}
    \label{fig:cost_quality}
\end{figure*}

\autoref{fig:cost_quality} shows the cost–quality trade-off directly, rather than combining cost and quality into a single ratio. The x-axis reports the number of tokens used per documentation entity, shown on a log scale because the methods differ by roughly an order of magnitude. The y-axis reports the documentation score. Thus, points closer to the upper-left are better, as they achieve higher scores with fewer tokens.
For visual compactness, each point uses the average score over the single-round and multi-round judge interactions. The separate S and M results are reported in \autoref{tab:rq2_extended}. For each method, the figure shows 10 small markers, one for each repository, together with a larger centroid that represents the method-level mean.

\ours{} sits in the upper-left region under both ISO and GitHub criteria---higher scores at roughly an order of magnitude lower token cost than Long-Context, and noticeably to the upper-left of RepoAgent. The decomposed view makes clear that the advantage comes from \emph{both} sides simultaneously: dependency-aware retrieval reduces redundant context (the x-axis), and standards-guided prompting raises the resulting documentation quality (the y-axis). \autoref{tab:token_iso_eval}  in the appendix reports absolute per-repository token usage for \ours{} under the ISO standard for reference, and \autoref{fig:evidence_redun} illustrates the same efficiency--quality combination qualitatively on a single function, where \ours{} matches RepoAgent's coverage of the key side-effects at substantially lower token cost.

\begin{table}[!t]
\centering
\caption{Component Ablation Across Ten Repositories Under the ISO Standard (Single-Round). Each Cell Shows the Average Score (1--5); Best per Repository in \textbf{Bold}.}
\label{tab:rq4_ablation_full}
\renewcommand{\arraystretch}{1.15}
\setlength{\tabcolsep}{6pt}
\footnotesize
\begin{tabular}{l|cccc}
\toprule
\textbf{Repository} & \textbf{Full} & \textbf{No Dep.} & \textbf{No Prompt} & \textbf{Baseline} \\
\midrule
NumPy    & \textbf{3.82} & 3.54 & 3.39 & 3.08 \\
OpenUI   & \textbf{3.58} & 3.32 & 3.21 & 2.89 \\
RAGFlow  & \textbf{3.74} & 3.48 & 3.35 & 3.02 \\
Dify     & \textbf{3.69} & 3.43 & 3.30 & 2.97 \\
Flask    & \textbf{3.78} & 3.51 & 3.42 & 3.11 \\
Pyro     & \textbf{3.61} & 3.37 & 3.25 & 2.91 \\
Fairseq  & \textbf{3.71} & 3.45 & 3.33 & 3.00 \\
Joblib   & \textbf{3.76} & 3.49 & 3.38 & 3.05 \\
Sonnet   & \textbf{3.63} & 3.38 & 3.27 & 2.93 \\
PyTorch  & \textbf{3.85} & 3.57 & 3.44 & 3.12 \\
\midrule
\textbf{Average} & \textbf{3.73} & 3.45 & 3.33 & 3.00 \\
\bottomrule
\end{tabular}
\end{table}

\begin{table}[!ht]
\centering
\caption{Static Dependency Recovery by \texttt{Pyan3} per Repository.}
\label{tab:rq6_dep_audit}
\renewcommand{\arraystretch}{1.15}
\setlength{\tabcolsep}{6pt}
\footnotesize
\begin{tabular}{l|ccc|c}
\toprule
\textbf{Repository} & \textbf{Precision} & \textbf{Recall} & \textbf{F1} & \textbf{Dynamic \%} \\
\midrule
Flask    & 91.2\% & 76.4\% & 83.1\% & 18.3\% \\
NumPy    & 88.7\% & 71.8\% & 79.4\% & 24.1\% \\
PyTorch  & 86.3\% & 67.2\% & 75.6\% & 31.5\% \\
\midrule
\textbf{Average} & 88.7\% & 71.8\% & 79.4\% & 24.6\% \\
\bottomrule
\end{tabular}
\end{table}

RQ3.b~--~Component ablation.
To attribute the quality gains, we isolate the two main design choices in \ours{}---dependency-aware retrieval and standards-guided structured prompting---via four variants:
\textbf{Full Method} (both components),
\textbf{No Dependency} (only the target entity, no dependency-relevant context),
\textbf{No Prompt} (dependency retrieval retained, but the structured prompt is replaced by a simplified prompt without standards-derived criteria), and
\textbf{Baseline} (both components removed).

\autoref{tab:rq4_ablation_full} reports the repository-wide ablation under the ISO single-round setting. The ordering Full~$>$ No~Dep.\ $>$ No~Prompt~$>$ Baseline holds in every repository. On average, removing dependency retrieval reduces the score by $0.28$ points, and removing the structured prompt reduces it by $0.40$ points relative to the full method---both components contribute consistently, with the standards-guided prompt being the larger driver of quality.

RQ3.c – Static dependency recovery audit. Because \ours{} relies on Pyan3 static call graphs for dependency-aware retrieval, the accuracy of recovered caller–callee edges directly affects the relevance of the injected context. We therefore audit Pyan3’s precision and recall on Flask, NumPy, and PyTorch, which span our repository-scale groups and vary in their use of dynamic Python features.
For each repository, we randomly sample 50 functions. Two authors independently annotate each function’s ground-truth callers and callees from the source code, resolving disagreements by discussion. We then compute precision as the fraction of recovered Pyan3 edges that are correct, and recall as the fraction of ground-truth edges that Pyan3 recovers. We also report the share of ground-truth edges involving dynamic features, such as \texttt{getattr}, parameterized decorators, metaclasses, or monkey-patching, since these are especially difficult for static analysis.

\autoref{tab:rq6_dep_audit} reports the audit. Precision is uniformly high (86--91\%): \texttt{Pyan3} rarely fabricates dependency edges. Recall is moderate (67--76\%), reflecting the inherent limitation of static analysis on dynamic-feature-heavy code. PyTorch exhibits the lowest recall and the highest dynamic-edge proportion (31.5\%). To quantify the downstream impact, we compare ISO scores for functions with complete dependency recovery (all ground-truth edges recovered) against those with at least one missed edge: $3.82$ vs.\ $3.51$ ($\Delta = 0.31$, Wilcoxon $p = 0.023$). Dependency completeness therefore correlates positively with downstream quality, indicating that future work on improving static-analysis coverage (or supplementing it with dynamic analysis) would directly translate into documentation gains.

\summary{\ours{} achieves higher documentation quality than both baselines at substantially lower token cost. The repository-wide ablation shows that both dependency-aware retrieval and standards-guided prompting contribute across all repositories, with the standards-guided prompt being the larger contributor. Static dependency recovery is high-precision but moderate-recall on dynamic-feature-heavy code; functions with complete recovery score $0.31$ points higher, identifying static-analysis coverage as a concrete leverage point for future improvement.
}

\section{Threats to Validity}
\label{sec:threats}

We organise threats to validity along construct, internal, and external dimensions.

\textit{Construct validity.}\quad Our evaluation relies on an LLM judge that scores documentation against predefined criteria, raising two concerns: judge reliability and rubric circularity. For reliability, RQ1.a shows that the judge responds consistently to controlled quality degradation, and RQ1.b corroborates it against human scores ($\rho = 0.72$, $\kappa = 0.51$) and across model families. For circularity---the same criteria drive both the generation prompt and the evaluation rubric---we note that the sensitivity test in RQ1.a uses official PyTorch documentation rather than \ours{} outputs, validating the rubric independently of our method, and that all baselines are scored under the same rubric so any criterion-alignment inflation affects them equally.

\textit{Internal validity.}\quad The Long-Context baseline isolates the contribution of dependency-aware retrieval from simply providing more code context, but additional baselines (e.g., retrieval-only approaches and other repository-context methods) would further strengthen the comparison. The relatively small repository sample ($n=10$) limits statistical power.
We therefore report per-repository results so consistency can be assessed directly. Generation, refinement, and judging in the default configuration all use the same model family (GPT-5-mini), introducing potential same-family bias; the cross-family comparison in RQ1.b provides partial mitigation. The criteria-extraction pipeline uses GPT-5-mini followed by two-author manual normalisation, with a full traceability table from source clauses to dimensions provided in the supplementary material; independent inter-annotator agreement remains future work.

\textit{External validity.}\quad The evaluation is limited to ten open-source Python repositories. \texttt{Pyan3}'s static analysis cannot fully capture dependencies arising from dynamic Python constructs (RQ3.c), and missed edges degrade downstream quality by $0.31$ points on average. Results may therefore not generalize to repositories that rely heavily on dynamic constructs or to other programming languages. 

\section{Related Work}
\label{sec:related_work}

Our work combines automatic code documentation generation, repository-level software understanding, and standards-guided evaluation. Unlike prior approaches that focus on method-level summarization or repository-level prompting, we ground both generation and evaluation in explicit criteria derived from authoritative standards~\cite{ISO}.

\subsection{Software Documentation and its Quality}

Software documentation is critical for software evolution, maintenance, and developer productivity, yet maintaining high-quality documentation remains challenging, especially in large and rapidly evolving systems~\cite{shafiq2021literature,zhong2015empirical,pietri2020determining,rai2022review}. While documentation quality depends on multiple dimensions, including accuracy, completeness, consistency, readability, and maintainability, these attributes are rarely defined through explicit and actionable criteria. We therefore treat documentation quality as an assessable engineering objective, grounding both generation and evaluation in established documentation standards rather than subjective judgments.

\subsection{Automatic Code Documentation Generation}

Early work on automatic code documentation generation relied on heuristic, template-based, and retrieval-oriented methods, using code stereotypes, method context, structural cues, and mined developer discussions to generate natural-language summaries for functions or code fragments~\cite{2025-LLM4CodeSum, 2024-EACS, abid2015using,mcburney2014automatic,rai2017method,vassallo2014codes,wong2015clocom,haiduc2010supporting}. Such approaches are interpretable but typically limited to local contexts, struggling to capture broader repository structure and cross-file dependencies. Neural models subsequently improved summarisation by learning code-to-text mappings~\cite{iyer2016summarizing,feng2020codebert,ahmad2021unified}, and the emergence of LLMs has enabled zero-shot and few-shot documentation generation with general-purpose LLMs~\cite{khan2022automatic,shin2021survey}. Much of this literature, however, still focuses on method- or snippet-level generation, where the target is relatively local and self-contained.

\subsection{Repository-Level Documentation Generation}

More recent work has moved beyond isolated functions toward repository-level documentation generation, since real repositories exhibit dependencies across files, modules, classes, and functions, and useful documentation often requires reasoning over this broader context. RepoAgent leverages repository structure and inter-file relationships to generate repository-level documentation~\cite{luo2024repoagent}, and RepoSummary explores feature-oriented summarisation that emphasises higher-level functional organisation over purely local descriptions~\cite{zhu2025reposummary}. These systems represent an important step toward practical documentation generation, but they generally define documentation quality \emph{implicitly}---through prompting strategies, output formats, or high-level user instructions---rather than grounding generation or evaluation in reusable criteria derived from standards or engineering guidelines. In contrast, our work makes documentation criteria first-class: fine-grained criteria are extracted from established standards and used to guide both generation and evaluation. Within the same per-entity scope, we additionally report cost--quality trade-offs and dependency-recovery audits that are typically absent from prior repository-level work.

\subsection{Evaluation of Documentation and Code Understanding Quality}

Evaluation remains a major challenge in automatic documentation generation. Earlier work relied on human judgement or small-scale comparative studies to assess summary usefulness and quality~\cite{eddy2013evaluating}, while broader software-engineering research emphasises the importance of rigorous and transparent evaluation methodology with careful treatment of datasets, criteria, and metrics~\cite{wang2021can}. Recent studies on code understandability similarly argue that meaningful evaluation should go beyond surface-level fluency and assess whether artifacts genuinely improve comprehension and developer usefulness~\cite{oliveira2021recommending,oliveira2024understanding}. The LLM-as-a-judge paradigm has emerged as a scalable alternative for evaluating generated text, with representative approaches including MT-Bench~\cite{zheng2023judging}, G-Eval~\cite{liu2023geval}, and JudgeLM~\cite{zhu2023judgelm}, showing that carefully designed rubrics and prompts can achieve reasonable agreement with human evaluators---a finding that motivates our standards-aligned judge design. In the broader context of repository-level code intelligence, retrieval-augmented approaches such as RepoCoder~\cite{zhang2024repocoder} and benchmarks such as RepoBench~\cite{liu2023repobench} demonstrate the value of selectively retrieving cross-file context for code completion; our dependency-aware retrieval shares this motivation but targets structurally relevant context for standards-guided documentation generation rather than completion candidates.

Our work differs from prior evaluation practice in two respects. First, rather than relying solely on unconstrained preference judgements, we construct an explicit rubric of documentation criteria derived from formal and practice-oriented standards. Second, we use this rubric to support criterion-level LLM-based evaluation, enabling repository documentation to be assessed in a structured and reproducible way that aligns documentation generation with broader software-engineering expectations.

\section{Conclusion and Future Work}

This paper presented \ours{}, a standards-guided and dependency-aware framework for code-level documentation generation. By aligning the generation target, the static-analysis input, and the evaluation unit on the same per-entity scope, and by extracting fine-grained criteria from formal ISO/IEC/IEEE standards and practice-oriented GitHub references, \ours{} grounds both generation and evaluation in a single reusable rubric, departing from prior systems that rely on implicit, ad-hoc prompts.

Our experiments show that the standards-derived criteria respond consistently to controlled degradation and that the LLM-based evaluation aligns with human judgement ($\rho = 0.72$) and across judge model families (RQ1); that \ours{} improves documentation quality over both RepoAgent and a long-context baseline, with significant effect sizes and broad per-category gains (RQ2); and that dependency-aware retrieval and standards-guided prompting contribute complementarily for a more favourable cost--quality trade-off, while static dependency recovery is high-precision but moderate-recall, with completeness correlating positively with downstream quality (RQ3).

Future work includes scaling the human evaluation, extending \ours{} beyond Python, supplementing static analysis with dynamic analysis to close the recall gap, and conducting independent inter-annotator validation of the extracted criteria. We believe the standards-guided perspective offers a useful foundation for documentation systems where quality must be assessed not only by fluency, but also by engineering completeness, practical usefulness, and maintainable organization.

\section*{Acknowledgement}
This work is partially supported by the National Key Research and Development Program of China (2024YFF0908000) and the National Natural Science Foundation of China (U24A20337 and 62372228) and 111 Center(No. B26023).

\bibliographystyle{IEEEtran}
\bibliography{refer}

@article{van2025life,
  title={The life of software features: An exploratory case study of 189 feature requests in Marlin},
  author={van der Hofstad, Aron and Cleophas, Loek and Dubslaff, Clemens and Kr{\"u}ger, Jacob},
  journal={Journal of Systems and Software},
  pages={112647},
  year={2025},
  publisher={Elsevier}
}

@misc{pep257,

  author       = {Goodger, David and van Rossum, Guido},

  title        = {{PEP 257 -- Docstring Conventions}},

  howpublished = {\url{https://peps.python.org/pep-0257/}},

  year         = {2001}

}

@misc{javadoc,

  author       = {{Oracle}},

  title        = {{Documentation Comment Specification for the Standard Doclet}},

  howpublished = {\url{https://docs.oracle.com/en/java/javase/24/docs/specs/javadoc/doc-comment-spec.html}},

  year         = {2025}

}

@misc{doxygen,

  author       = {{Doxygen}},

  title        = {{Documenting the Code}},

  howpublished = {\url{https://www.doxygen.nl/manual/docblocks.html}}

}

@inproceedings{barone2017parallel,

  title     = {{A Parallel Corpus of Python Functions and Documentation Strings for Automated Code Documentation and Code Generation}},

  author    = {Miceli Barone, Antonio Valerio and Sennrich, Rico},

  booktitle = {Proceedings of the Eighth International Joint Conference on Natural Language Processing},

  year      = {2017},

  pages     = {314--319}

}

@inproceedings{2025-LLM4CodeSum,
	address = {Ottawa, Ontario, Canada},
	author = {Weisong Sun and Yun Miao and Yuekang Li and Hongyu Zhang and Chunrong Fang and Yi Liu and Gelei Deng and Yang Liu and Zhenyu Chen},
	booktitle = {Proceedings of the 47th IEEE/ACM International Conference on Software Engineering},
	month = {27 April-3 May},
	pages = {419--431},
	title = {Source Code Summarization in the Era of Large Language Models},
	year = {2025}}

@article{2024-EACS,
	author = {Weisong Sun and Chunrong Fang and Yuchen Chen and Quanjun Zhang and Guanhong Tao and Yudu You and Tingxu Han and Yifei Ge and Yuling Hu and Bin Luo and Zhenyu Chen},
	journal = {ACM Transactions on Software Engineering and Methodology},
	number = {3},
	pages = {1--39},
	title = {An Extractive-and-Abstractive Framework for Source Code Summarization},
	volume = {33},
	year = {2024}}

@inproceedings{feng2020codebert,

  title     = {{CodeBERT: A Pre-Trained Model for Programming and Natural Languages}},

  author    = {Feng, Zhangyin and Guo, Daya and Tang, Duyu and Duan, Nan and Feng, Xiaocheng and Gong, Ming and Shou, Linjun and Qin, Bing and Liu, Ting and Jiang, Daxin and Zhou, Ming},
  booktitle = {Findings of EMNLP},
  year      = {2020},
  pages     = {1536--1547}
}

@standard{IEEE8292008,
  title        = {{IEEE Standard for Software and System Test Documentation}},
  organization = {{IEEE}},
  institution  = {{Institute of Electrical and Electronics Engineers}},
  number       = {{IEEE Std 829-2008}},
  year         = {2008},
  doi          = {10.1109/IEEESTD.2008.4578383}
}

@article{wang2021can,
  title={Can we benchmark code review studies? a systematic mapping study of methodology, dataset, and metric},
  author={Wang, Dong and Ueda, Yuki and Kula, Raula Gaikovina and Ishio, Takashi and Matsumoto, Kenichi},
  journal={Journal of Systems and Software},
  volume={180},
  pages={111009},
  year={2021},
  publisher={Elsevier}
}

@incollection{rajlich2014software,
  title={Software evolution and maintenance},
  author={Rajlich, V{\'a}clav},
  booktitle={Future of Software Engineering Proceedings},
  pages={133--144},
  year={2014}
}

@inproceedings{zhong2015empirical,
  title={An empirical study on real bug fixes},
  author={Zhong, Hao and Su, Zhendong},
  booktitle={37th IEEE International Conference on Software Engineering},
  volume={1},
  pages={913--923},
  year={2015},
  organization={IEEE}
}

@inproceedings{pietri2020determining,
  title={Determining the intrinsic structure of public software development history},
  author={Pietri, Antoine and Rousseau, Guillaume and Zacchiroli, Stefano},
  booktitle={Proceedings of the 17th international conference on mining software repositories},
  pages={602--605},
  year={2020}
}

@article{ruparelia2010software,
  title={Software development lifecycle models},
  author={Ruparelia, Nayan B},
  journal={ACM SIGSOFT Software Engineering Notes},
  volume={35},
  number={3},
  pages={8--13},
  year={2010},
  publisher={ACM New York, NY, USA}
}

@article{georgiou2019software,
  title={Software development lifecycle for energy efficiency: techniques and tools},
  author={Georgiou, Stefanos and Rizou, Stamatia and Spinellis, Diomidis},
  journal={ACM Computing Surveys},
  volume={52},
  number={4},
  pages={1--33},
  year={2019},
  publisher={ACM New York, NY, USA}
}

@article{shafiq2021literature,
  title={A literature review of using machine learning in software development life cycle stages},
  author={Shafiq, Saad and Mashkoor, Atif and Mayr-Dorn, Christoph and Egyed, Alexander},
  journal={IEEE Access},
  volume={9},
  pages={140896--140920},
  year={2021},
  publisher={IEEE}
}

@article{oliveira2024understanding,
  title={Understanding Code Understandability Improvements in Code Reviews},
  author={Oliveira, Delano and Santos, Reydne and De Oliveira, Benedito and Monperrus, Martin and Castor, Fernando and Madeiral, Fernanda},
  journal={IEEE Transactions on Software Engineering},
  year={2024},
  publisher={IEEE}
}

@inproceedings{oliveira2021recommending,
  title={Recommending code understandability improvements based on code reviews},
  author={Oliveira, Delano},
  booktitle={36th IEEE/ACM International Conference on Automated Software Engineering Workshops},
  pages={131--132},
  year={2021},
  organization={IEEE}
}

@article{rai2022review,
  title={A review on source code documentation},
  author={Rai, Sawan and Belwal, Ramesh Chandra and Gupta, Atul},
  journal={ACM Transactions on Intelligent Systems and Technology},
  volume={13},
  number={5},
  pages={1--44},
  year={2022},
  publisher={ACM New York, NY}
}

@article{alzahrani2024software,
  title={Software Systems Documentation: A Systematic Review.},
  author={Alzahrani, Abdullah AH},
  journal={International Journal of Advanced Computer Science \& Applications},
  volume={15},
  number={8},
  year={2024}
}

@article{shin2021survey,
  title={A survey of automatic code generation from natural language},
  author={Shin, Jiho and Nam, Jaechang and others},
  journal={Journal of Information Processing Systems},
  volume={17},
  number={3},
  pages={537--555},
  year={2021},
  publisher={Korea Information Processing Society}
}

@article{zhu2025reposummary,
  title={RepoSummary: Feature-Oriented Summarization and Documentation Generation for Code Repositories},
  author={Zhu, Yifeng and Zhao, Xianlin and Li, Xutian and Zou, Yanzhen and Yuan, Haizhuo and Wang, Yue and Xie, Bing},
  journal={arXiv preprint arXiv:2510.11039},
  year={2025}
}

@inproceedings{khan2022automatic,
  title={Automatic code documentation generation using gpt-3},
  author={Khan, Junaed Younus and Uddin, Gias},
  booktitle={Proceedings of the 37th IEEE/ACM International Conference on Automated Software Engineering},
  pages={1--6},
  year={2022}
}

@inproceedings{luo2024repoagent,
  title={RepoAgent: An LLM-Powered Open-Source Framework for Repository-level Code Documentation Generation},
  author={Luo, Qinyu and Ye, Yining and Liang, Shihao and Zhang, Zhong and Qin, Yujia and Lu, Yaxi and Wu, Yesai and Cong, Xin and Lin, Yankai and Zhang, Yingli and others},
  booktitle={2024 Conference on Empirical Methods in Natural Language Processing: System Demonstrations},
  pages={436--464},
  year={2024}
}

@inproceedings{abid2015using,
  title={Using stereotypes in the automatic generation of natural language summaries for c++ methods},
  author={Abid, Nahla J and Dragan, Natalia and Collard, Michael L and Maletic, Jonathan I},
  booktitle={IEEE International Conference on Software Maintenance and Evolution},
  pages={561--565},
  year={2015},
  organization={IEEE}
}

@inproceedings{mcburney2014automatic,
  title={Automatic documentation generation via source code summarization of method context},
  author={McBurney, Paul W and McMillan, Collin},
  booktitle={Proceedings of the 22nd International Conference on Program Comprehension},
  pages={279--290},
  year={2014}
}

@inproceedings{rai2017method,
  title={Method level text summarization for java code using nano-patterns},
  author={Rai, Sawan and Gaikwad, Tejaswini and Jain, Sparshi and Gupta, Atul},
  booktitle={24th Asia-Pacific Software Engineering Conference},
  pages={199--208},
  year={2017},
  organization={IEEE}
}

@inproceedings{eddy2013evaluating,
  title={Evaluating source code summarization techniques: Replication and expansion},
  author={Eddy, Brian P and Robinson, Jeffrey A and Kraft, Nicholas A and Carver, Jeffrey C},
  booktitle={International Conference on Program Comprehension},
  pages={13--22},
  year={2013},
  organization={IEEE}
}

@inproceedings{vassallo2014codes,
  title={Codes: Mining source code descriptions from developers discussions},
  author={Vassallo, Carmine and Panichella, Sebastiano and Di Penta, Massimiliano and Canfora, Gerardo},
  booktitle={Proceedings of the 22nd International Conference on Program Comprehension},
  pages={106--109},
  year={2014}
}

@inproceedings{wong2015clocom,
  title={Clocom: Mining existing source code for automatic comment generation},
  author={Wong, Edmund and Liu, Taiyue and Tan, Lin},
  booktitle={22nd International conference on software analysis, evolution, and reengineering},
  pages={380--389},
  year={2015},
  organization={IEEE}
}

@inproceedings{haiduc2010use,
  title={On the use of automated text summarization techniques for summarizing source code},
  author={Haiduc, Sonia and Aponte, Jairo and Moreno, Laura and Marcus, Andrian},
  booktitle={2010 17th Working conference on reverse engineering},
  pages={35--44},
  year={2010},
  organization={IEEE}
}

@inproceedings{haiduc2010supporting,
  title={Supporting program comprehension with source code summarization},
  author={Haiduc, Sonia and Aponte, Jairo and Marcus, Andrian},
  booktitle={Proceedings of the 32nd ACM/IEEE International Conference on Software Engineering-Volume 2},
  pages={223--226},
  year={2010}
}

@inproceedings{iyer2016summarizing,
  title={Summarizing source code using a neural attention model},
  author={Iyer, Srinivasan and Konstas, Ioannis and Cheung, Alvin and Zettlemoyer, Luke},
  booktitle={54th Annual Meeting of the Association for Computational Linguistics},
  pages={2073--2083},
  year={2016},
  organization={Association for Computational Linguistics}
}

@techreport{ISO,
  title={{ISO/IEC/IEEE 26514:2018 --- Systems and software engineering --- Developing information for users in an agile environment}},
  author={{ISO/IEC/IEEE}},
  institution={International Organization for Standardization},
  year={2018},
  type={International Standard}
}

@misc{iso12207,
  author       = {{ISO/IEC/IEEE}},
  title        = {{ISO/IEC/IEEE} 12207:2017 --- Systems and Software Engineering --- Software Life Cycle Processes},
  howpublished = {International Standard},
  year         = {2017},
}

@inproceedings{ahmad2021unified,
  title={Unified Pre-training for Program Understanding and Generation},
  author={Ahmad, Wasi Uddin and Chakraborty, Saikat and Ray, Baishakhi and Chang, Kai-Wei},
  booktitle={Proceedings of the 2021 Conference of the North American Chapter of the Association for Computational Linguistics: Human Language Technologies},
  pages={2655--2668},
  year={2021}
}

@misc{pytorch_docs,
  author       = {{PyTorch Contributors}},
  title        = {{PyTorch} Documentation},
  howpublished = {\url{https://pytorch.org/docs/2.8/}},
  note         = {Version 2.8},
  year         = {2025},
}

@inproceedings{zheng2023judging,
  title={Judging {LLM}-as-a-Judge with {MT}-Bench and Chatbot Arena},
  author={Zheng, Lianmin and Chiang, Wei-Lin and Sheng, Ying and Zhuang, Siyuan and Wu, Zhanghao and Zhuang, Yonghao and Lin, Zi and Li, Zhuohan and Li, Dacheng and Xing, Eric P. and others},
  booktitle={Advances in Neural Information Processing Systems},
  volume={36},
  year={2023}
}

@inproceedings{liu2023geval,
  title={{G-Eval}: {NLG} Evaluation using {GPT-4} with Better Human Alignment},
  author={Liu, Yang and Iter, Dan and Xu, Yichong and Wang, Shuohang and Xu, Ruochen and Zhu, Chenguang},
  booktitle={Proceedings of the 2023 Conference on Empirical Methods in Natural Language Processing},
  pages={2511--2522},
  year={2023}
}

@inproceedings{zhang2024repocoder,
  title={{RepoCoder}: Repository-Level Code Completion Through Iterative Retrieval and Generation},
  author={Zhang, Fengji and Chen, Bei and Zhang, Yue and Keung, Jacky and Liu, Jin and Zan, Daoguang and Mao, Yi and Lou, Jian-Guang and Chen, Weizhu},
  booktitle={Proceedings of the 2023 Conference on Empirical Methods in Natural Language Processing},
  pages={2471--2484},
  year={2023}
}

@inproceedings{liu2023repobench,
  title={{RepoBench}: Benchmarking Repository-Level Code Auto-Completion Systems},
  author={Liu, Tianyang and Xu, Canwen and McAuley, Julian},
  booktitle={The Twelfth International Conference on Learning Representations},
  year={2024}
}

@inproceedings{zhu2023judgelm,
  title={{JudgeLM}: Fine-tuned Large Language Models are Scalable Judges},
  author={Zhu, Lianghui and Wang, Xinggang and Wang, Xinlong},
  booktitle={arXiv preprint arXiv:2310.17631},
  year={2023}
}

\clearpage
\appendix

\begin{table*}[!t]
\centering
\caption{Statistics of the ten evaluated open-source Python repositories. For each repository, we report the repository URL, the numbers of extracted classes (\#Class) and functions (\#Function), and the resulting total number of documentation entities used as the structural complexity measure. Following this measure, repositories are grouped into three scale ranges: small, medium, and large.}
\label{tab:repo_stats}
\renewcommand{\arraystretch}{1.2}
\setlength{\tabcolsep}{6pt}
\footnotesize
\begin{tabular}{l l r r r l}
\toprule
\textbf{Name} & \textbf{Repository URL} & \textbf{\#Class} & \textbf{\#Function} & \textbf{Total} & \textbf{Scale} \\
\midrule
NumPy     & \url{https://github.com/numpy/numpy}                  & 5   & 31   & 36   & \multirow{3}{*}{Small} \\
OpenUI    & \url{https://github.com/wandb/openui}                 & 3   & 21   & 24   &  \\
RAGFlow   & \url{https://github.com/infiniflow/ragflow}           & 1   & 14   & 15   &  \\
\midrule
Dify      & \url{https://github.com/langgenius/dify}              & 3   & 47   & 50   & \multirow{3}{*}{Medium} \\
Flask     & \url{https://github.com/pallets/flask}                & 28  & 207  & 235  &  \\
Pyro      & \url{https://github.com/pyro-ppl/pyro}                & 28  & 234  & 262  &  \\
\midrule
Fairseq   & \url{https://github.com/facebookresearch/fairseq}     & 45  & 378  & 423  & \multirow{4}{*}{Large} \\
Joblib    & \url{https://github.com/joblib/joblib}                & 56  & 389  & 445  &  \\
Sonnet    & \url{https://github.com/google-deepmind/sonnet}       & 177 & 801  & 978  &  \\
PyTorch   & \url{https://github.com/pytorch/pytorch}              & 71  & 1279 & 1350 &  \\
\bottomrule
\end{tabular}
\end{table*}

\begin{table}[!t]
\centering
\caption{Per-repository token usage of \ours{} under the ISO standard, reported separately for input and output (cumulative across the generation pipeline). Entity counts per repository are given in \autoref{tab:repo_stats}.}
\label{tab:token_iso_eval}
\renewcommand{\arraystretch}{1.15}
\setlength{\tabcolsep}{6pt}
\footnotesize
\begin{tabular}{l|rrr}
\toprule
\textbf{Repository} & \textbf{Input} & \textbf{Output} & \textbf{Total} \\
\midrule
numpy-2.3.4    & 46,004    & 21,865  & 67,869 \\
openui         & 32,614    & 11,057  & 43,671 \\
ragflow-0.21.1 & 21,327    & 18,235  & 39,562 \\
dify-1.9.1     & 64,604    & 27,999  & 92,603 \\
flask-3.1.2    & 308,876   & 122,794 & 431,670 \\
pyro-1.9.1     & 464,061   & 168,188 & 632,249 \\
fairseq-0.12.2 & 628,851   & 268,095 & 896,946 \\
joblib-1.5.2   & 620,872   & 260,111 & 880,983 \\
sonnet-2.0.2   & 1,196,715 & 511,800 & 1,708,515 \\
pytorch-2.8    & 2,160,693 & 913,178 & 3,073,871 \\
\bottomrule
\end{tabular}
\end{table}

\begin{figure*}[!t]
    \centering
    \includegraphics[width=0.82\linewidth]{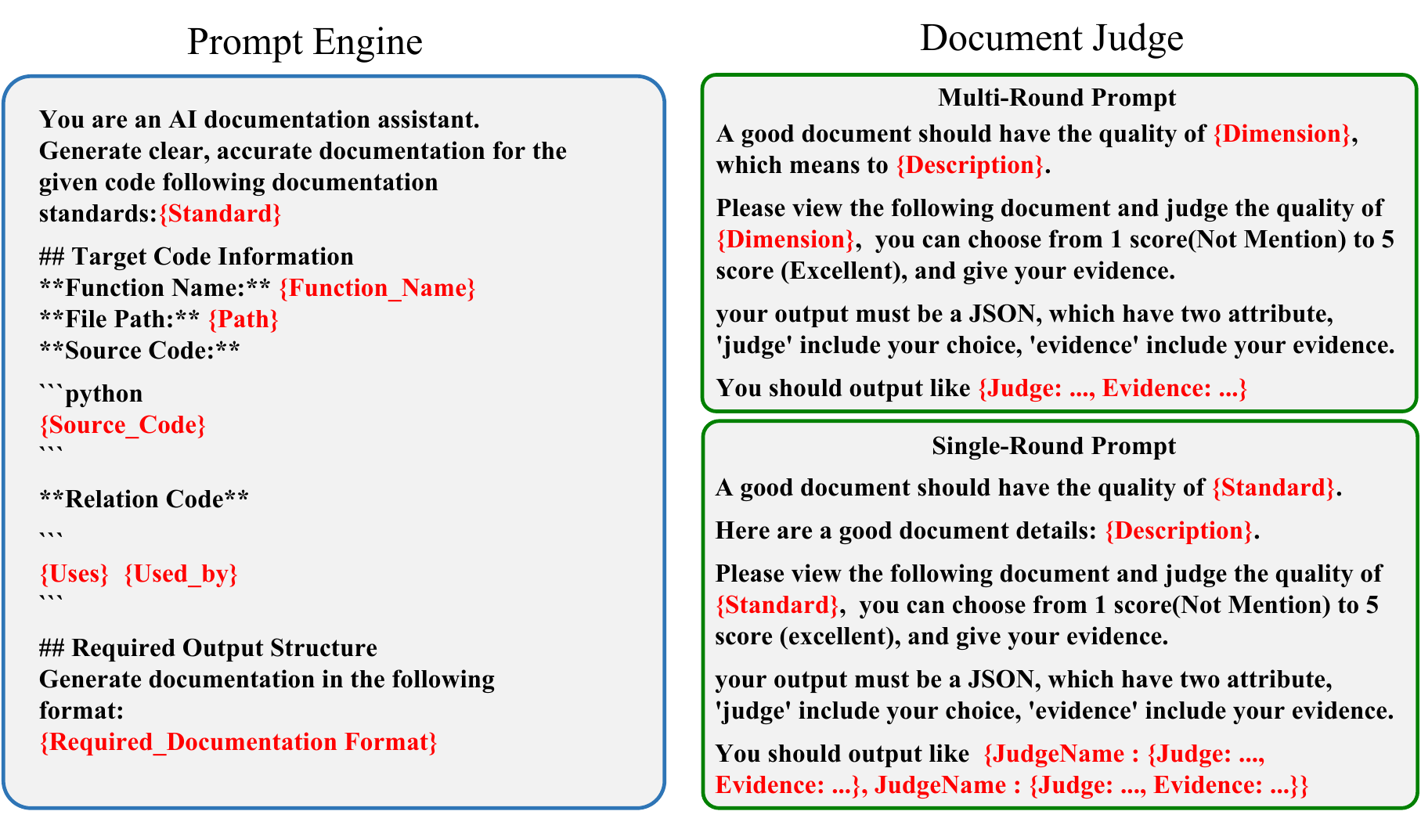}
    \caption{The prompt used for our components ``Prompt Engine'' and ``Document Judge'' in \autoref{fig:framework}.}
    \vspace{-8pt}
    \label{fig:prompt}
\end{figure*}

\subsection{Evaluation Rubrics}
\label{app:rubrics}
This appendix lists the complete criterion set $\mathcal{Q}$ produced by the offline extraction in Phase~(a).
\autoref{tab:iso_dimensions} gives the ISO rubric (7 categories, 14 dimensions), derived from ISO/IEC/IEEE~26514, and \autoref{tab:github_dimensions} gives the GitHub rubric (5 categories, 12 dimensions), derived from the GitHub documentation style guide. For each dimension, we report the short name used as the judge variable and the one-sentence definition supplied to the judge; the 1--5 scoring anchors, and the full clause-to-dimension traceability table.
\begin{table*}[!t]
\centering
\caption{ISO-based evaluation rubric extracted from the ISO/IEC/IEEE software documentation standard. The table summarizes the evaluation dimensions used in our experiments, together with brief descriptions of each dimension for criterion-aligned LLM-based documentation assessment.}
\label{tab:iso_dimensions}
\renewcommand{\arraystretch}{1.2}
\setlength{\tabcolsep}{6pt}
\footnotesize
\begin{tabular}{p{2.6cm} p{3.3cm} p{11.2cm}}
\toprule
\textbf{Category} & \textbf{Dimension} & \textbf{Description} \\
\midrule

\multirow{2}{*}{\parbox[t]{2.6cm}{Content Quality}}
& Conciseness & Avoid redundancy and repetition, and include only necessary information. \\
\cmidrule(lr){2-3}
& Legibility & Font size and color contrast should be suitable for the reading environment. \\
\midrule

\multirow{3}{*}{\parbox[t]{2.6cm}{Information Structure}}
& Modularity & Content should be divided into independent, reusable topics. \\
\cmidrule(lr){2-3}
& Navigability & Provide table of contents, index, and search functionality. \\
\cmidrule(lr){2-3}
& Metadata & Each topic should include metadata for content management and retrieval. \\
\midrule

\multirow{2}{*}{\parbox[t]{2.6cm}{Format \& Presentation}}
& Consistent Format & Use unified style guides, including fonts, colors, icons, and warning formats. \\
\cmidrule(lr){2-3}
& Interactive Element Representation & Use consistent graphics or formats to represent keyboard shortcuts. \\
\midrule

\multirow{3}{*}{\parbox[t]{2.6cm}{i18n \& Localization}}
& Cultural Neutrality & Avoid culture-specific examples, humor, or images. \\
\cmidrule(lr){2-3}
& Translation Friendly & Use simple sentences and consistent terminology. \\
\cmidrule(lr){2-3}
& Space Consideration & Reserve space for text expansion in different languages. \\
\midrule

\parbox[t]{2.6cm}{Usability \& UX}
& Search and Indexing & Provide full-text search, Boolean search, and filtering functions. \\
\midrule

\multirow{2}{*}{\parbox[t]{2.6cm}{Safety \& Compliance}}
& Safety Information & Danger, warning, and caution information should be prominent. \\
\cmidrule(lr){2-3}
& Legal and Copyright & Include copyright notices, trademark information, and license terms. \\
\midrule

Maintenance \& Updates
& Change Log & Clearly identify new or modified content, and provide update notes. \\
\bottomrule
\end{tabular}
\end{table*}

\begin{table*}[!t]
\centering
\caption{GitHub-based evaluation rubric extracted from the GitHub documentation style guide. The table summarizes the evaluation dimensions used in our experiments, together with brief descriptions of each dimension for criterion-aligned LLM-based documentation assessment.}
\label{tab:github_dimensions}
\renewcommand{\arraystretch}{1.2}
\setlength{\tabcolsep}{6pt}
\footnotesize
\begin{tabular}{p{2.6cm} p{3.3cm} p{11.2cm}}
\toprule
\textbf{Category} & \textbf{Dimension} & \textbf{Description} \\
\midrule

\multirow{2}{*}{\parbox[t]{2.6cm}{Content Model}}
& Title Standards & Ensure titles are clear, descriptive, and concise, typically within 60--80 characters. \\
\cmidrule(lr){2-3}
& Category Title Naming & Use task-based gerund titles that remain general and scalable. \\
\midrule

\multirow{4}{*}{\parbox[t]{2.6cm}{Article Content}}
& Permission Declaration Standards & State permissions upfront using the form ``People with [role] can [action].'' \\
\cmidrule(lr){2-3}
& Process Content Standards & Present procedures as clear step-by-step instructions. \\
\cmidrule(lr){2-3}
& Next Steps Standards & Include follow-up links that guide users to the next relevant topic. \\
\cmidrule(lr){2-3}
& Prerequisites Standards & List warnings and requirements before sensitive tasks. \\
\midrule

\multirow{3}{*}{\parbox[t]{2.6cm}{Reference Content}}
& Reference Content & Decide whether reference information should be presented as standalone articles or embedded sections. \\
\cmidrule(lr){2-3}
& Conceptual Content Guidelines & Begin with ``About [topic]'' and explain purpose and use cases clearly. \\
\cmidrule(lr){2-3}
& Error Reference Content Guidelines & Help users recognize and understand issues, separate from troubleshooting. \\
\midrule

\multirow{2}{*}{\parbox[t]{2.6cm}{Tutorial Content}}
& Tutorial Content Guidelines & Define the target audience and goals, and use a clear developer-oriented tone with examples. \\
\cmidrule(lr){2-3}
& Getting Started Content Guidelines & Provide a concise step-based quick start, preferably under 600 words. \\
\midrule

Further Reading
& Further Reading Standards & Provide a bulleted list of external or advanced resources. \\
\bottomrule
\end{tabular}
\end{table*}

\subsection{Evaluated Repositories}
\label{app:repos}
\autoref{tab:repo_stats} lists the ten evaluated repositories with their URLs,
the number of extracted classes and functions, the resulting
documentation-entity totals used as the structural-complexity measure, and the
small/medium/large scale grouping.

\subsection{Per-Repository Token Usage}
\label{app:tokens}
\autoref{tab:token_iso_eval} reports the absolute per-repository token consumption
of \ours{} under the ISO standard, separated into input and output tokens and
cumulative across the generation pipeline. The per-entity values plotted on the
$x$-axis of \autoref{fig:cost_quality} are obtained by dividing these totals
by the entity counts in \autoref{tab:repo_stats}.

\vfill

\end{document}